\documentclass[prb,twocolumn,showpacs,superscriptaddress,amsmath,amssymb]{revtex4}

\usepackage{graphicx}
\usepackage{dcolumn}
\usepackage{bm}
\usepackage[dvips]{color}

\newcommand{\vect}[1] {\textbf{{\textit {#1}}}}
\newcommand{\figwidth}{0.98\columnwidth}

\newcommand{\vectmu}{\boldsymbol\mu}
\begin{document}

\title{ Magnetic phase diagram of the frustrated $S=1/2$ chain magnet LiCu$_2$O$_2$}

\author{A.A. Bush}
\affiliation{Moscow Institute of Radioengineering, Electronics and
Automation, 117454 Moscow, Russia}

\author{V.N. Glazkov}
\affiliation{P. L. Kapitza Institute for Physical Problems RAS,
119334 Moscow, Russia}

\affiliation{Neutron Scattering and Magnetism, Laboratory for Solid State
Physics, ETH Zurich, Switzerland,  8093 Z\"{u}rich, Switzerland}

\author{M. Hagiwara}
\affiliation{KYOKUGEN, Osaka University, Machikaneyama 1-3, Toyanaka
560-8531, Japan}

\author{T. Kashiwagi}
\affiliation{Institute for Materials Science and Graduate School of Pure and
Applied Sciences, University of Tsukuba, 1-1-1 Tennodai, Tsukuba, Ibaraki
305-8573, Japan}

\author{S. Kimura}
\affiliation{Institute for Materials Research, Tohoku University, 2-1-1
Katahira, Sendai, Miyagi 980-8577, Japan}

\author{K.~Omura}
\affiliation{KYOKUGEN, Osaka University, Machikaneyama 1-3, Toyanaka
560-8531, Japan}

\author{L.A. Prozorova}
\affiliation{P. L. Kapitza Institute for Physical Problems RAS,
119334 Moscow, Russia}

\author{L.E. Svistov}

\email{svistov@kapitza.ras.ru}

\affiliation{P. L. Kapitza Institute for Physical Problems RAS,
119334 Moscow, Russia}

\affiliation{KYOKUGEN, Osaka University, Machikaneyama 1-3, Toyanaka
560-8531, Japan}

\author{A.M. Vasiliev}
\affiliation{P. L. Kapitza Institute for Physical Problems RAS,
119334 Moscow, Russia}

\author{A. Zheludev}
\affiliation{Neutron Scattering and Magnetism, Laboratory for Solid State
Physics, ETH Zurich, Switzerland,  8093 Z\"{u}rich, Switzerland}

\date{\today}

\begin{abstract}
We present the results of the magnetization and dielectric constant
measurements on untwinned single crystal samples of the frustrated $S=1/2$
chain cuprate LiCu$_2$O$_2$. Novel magnetic phase transitions were observed. A
spin flop transition of the spiral spin plane was observed for the field
orientations $\vect{H}
\parallel \vect{a}$, $\vect{b}$.   The second magnetic transition was observed at
$H \approx$~15 T for all three principal field directions. This high field
magnetic phase is discussed as a collinear spin-modulated phase which is
expected for an $S$=1/2 nearest-neighbor ferromagnetic and
next-nearest-neighbor antiferromagnetic chain system.

\end{abstract}

\maketitle

\section{Introduction}
Unconventional magnetic orders and phases in frustrated quantum--spin chains
are one of the attractive issues, because they appear under a fine balance of
the exchange interactions and are sometimes caused by much weaker interactions
or fluctuations~\cite{Chubukov_1991,Kolezhuk_2000,Kolezhuk_2005, Vekua_2007,
Dmitriev_2008}.

A  kind of frustration in quasi--one--dimensional (1D) chain magnets is
provided by competing interactions when the nearest neighbor (NN) exchange is
ferromagnetic and the next--nearest neighbor (NNN) exchange is
antiferromagnetic. Numerical investigations of the frustrated chain magnets
within different models~\cite{Heidrich_2006, Hikihara_2008, Sudan_2009,
Heidrich_2009} have predicted a number of exotic magnetic phases in the
magnetization process such as planar spiral and different multipolar phases.
Theoretical studies of the magnetic phase diagram show that the magnetic phases
are very sensitive to the interchain interactions and anisotropic interactions
in the system~\cite{Furukawa_2008,
Furukawa_2010,Zhitomirsky_2010,Nishimoto_2010,Sato_2011}.

 There are a number of magnets which are attractive objects for the
experimental investigations as realizations of one-dimensional frustrated
systems: LiCuVO$_4$, Rb$_2$Cu$_2$Mo$_3$O$_{12}$, Li$_2$ZrCuO$_4$, CuCl$_2$,
Li$_2$CuO$_2$, LiCu$_2$O$_2$, NaCu$_2$O$_2$ (see for example,
Refs.\onlinecite{Enderle_2005, Hase_2004, Drechsler_2007, Banks_2007,
Masuda_2005}). Some of these magnets, such as LiCuVO$_4$, Li$_2$CuO$_2$ and
Li$_2$ZrCuO$_4$, are quasi--one--dimensional magnets, while others are
quasi--two--dimensional magnets with the layers of frustrated chains.

In the present paper, we report the results of magnetization and dielectric
constant measurements on the quasi two dimensional antiferromagnet
LiCu$_2$O$_2$. Our new experiments expand the magnetic field range from the
fields of up to $9$~T of the previous work~\cite{Svistov_2009} to the fields of
up to $52$~T, which is approximately $40$\% of the expected saturation field
($H_{sat}$). These measurements display two phase transitions. The low field
transition may be interpreted as a spin-flop transition while the novel high
field transition is, most probably, of the exchange origin. Using the results
of a 1D frustrated model, we suggest that the observed high field transition
from a spiral magnetic phase is the transition to a collinear spin-modulated
magnetic phase.

\begin{figure}
\includegraphics[width=\figwidth,angle=0,clip]{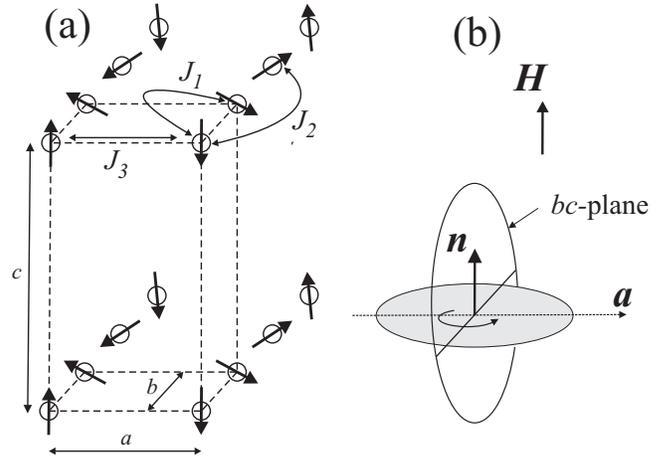}
\caption{(a) Schematic representation of the arrangement of Cu$^{2+}$ moments
in LiCu$_2$O$_2$. Only one of four Cu$^{2+}$ ($\alpha$, $\beta$, $\gamma$,
$\delta$) positions is shown. Arrows correspond to the spin directions below
$T_{c2}$. $J_1, J_2$ and $J_3$ are the main exchange
integrals~\cite{Masuda_2005}. (b) Scheme of the orientation of the spin plane
for $\vect{H}$ in the $(bc)$--plane. $H > H_{c1}$. Spin plane is shadowed with
gray. The direction of spin rotation by displacement along the chain
($b$--axes) is shown by the circular arrow. $\vect{n}$ is the vector normal to
the spin plane. } \label{fig:spins}
\end{figure}

\begin{figure}
\includegraphics[width=\figwidth,angle=0,clip]{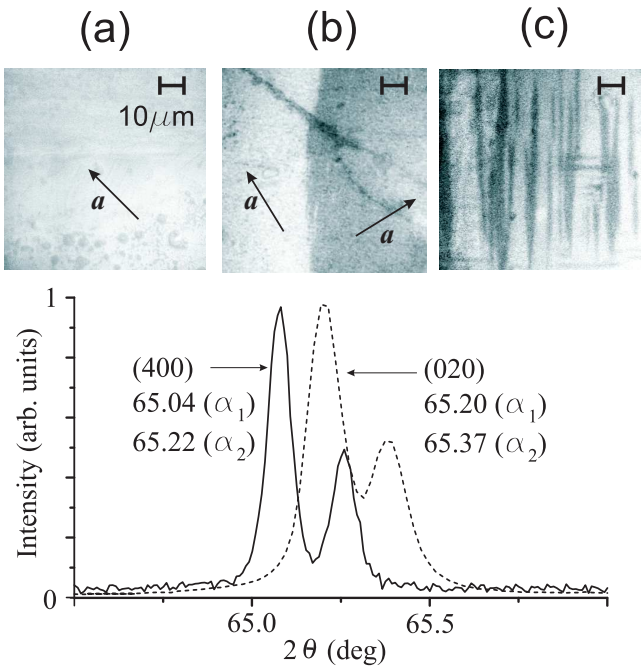}
\caption{Upper panel:Polarized optical microscope images of the $(ab)$--plane
of LiCu$_2$O$_2$ samples (a) without twinning, (b) with two large twinned
domains and (c) with complicated twinning structure. Lower panel: X-ray
diffraction pattern observed in $\theta-2\theta$ geometry on untwinned single
crystal. The spectra were obtained from the $(bc)$ and the $(ac)$--planes of
the crystal. } \label{fig:domains}
\end{figure}

\begin{figure}
\includegraphics[width=\figwidth,angle=0,clip]{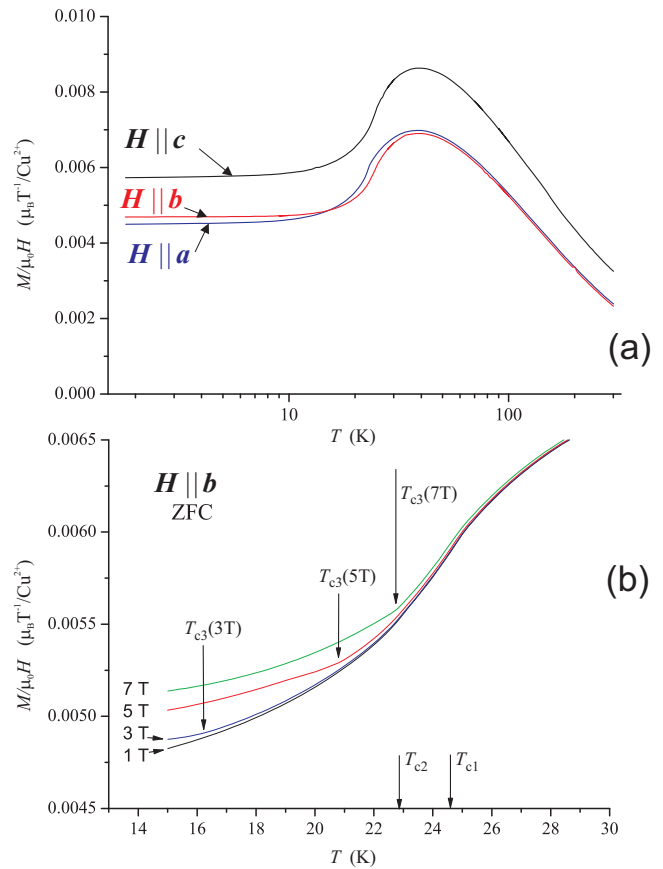}
\caption{(color online) Upper panel: Temperature dependence of the low field
magnetization curves $M(T)/\mu_0 H$ for three field directions $\vect{H}
\parallel \vect{a}, \vect{b}, \vect{c}$ and $\mu_0 H=0.1$~T.  Lower panel: $M(T)/\mu_0 H$ in the
vicinity of the transition temperatures $T_{c1}$ and $T_{c2}$ at $\mu_0 H =$1,
3, 5, and 7~T for $\vect{H} \parallel \vect{b}$.} \label{fig:M(T)}
\end{figure}

\begin{figure}
\includegraphics[width=\figwidth,angle=0,clip]{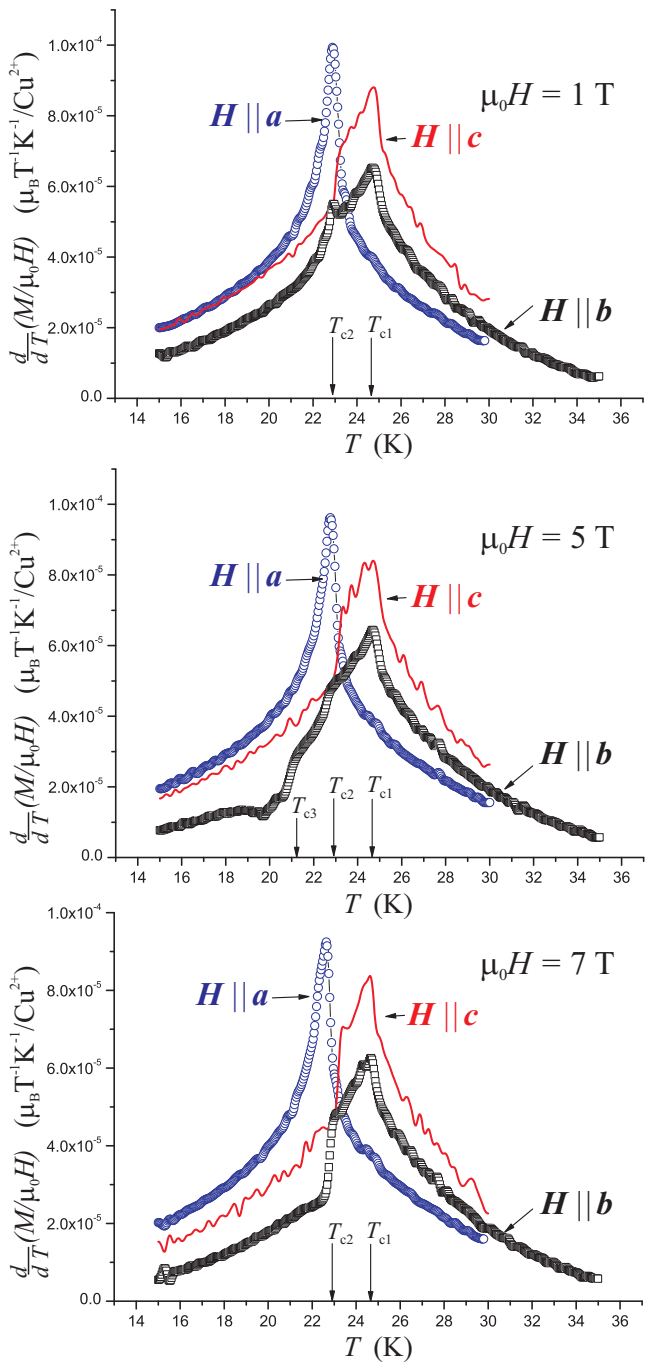}
\caption{(color online) Temperature dependences of the magnetization derivative
$dM/dT$ for three field directions in the vicinity of $T_{c1}$ and $T_{c2}$
obtained from $M(T)$ data at $\mu_{0}H=$1, 5, and 7~T.  } \label{fig:dmdt}
\end{figure}

\section{Crystallographic and magnetic structures of LiCu$_2$O$_2$}
LiCu$_2$O$_2$ crystallizes in the orthorhombic lattice (space group $Pnma$)
with the unit cell parameters $a=$5.73 \AA, $b=$2.86 \AA \, and $c=$12.42
\AA.\cite{Berg_1991} The unit cell parameter $a$ is approximately twice the
unit cell parameter $b$. Consequently, the LiCu$_2$O$_2$ samples, as a rule,
are characterized by the twinning due to formation of the crystallographic
domains rotated by $90^\circ$ around their common crystallographic axis $c$.

 The unit cell of the LiCu$_2$O$_2$ crystal contains
four univalent nonmagnetic cations Cu$^+$ and four divalent cations Cu$^{2+}$
with the spin $S$=1/2. There are four equivalent crystallographic positions of
the magnetic Cu$^{2+}$ ions in the crystal unit cell of LiCu$_2$O$_2$,
conventionally denoted as $\alpha$, $\beta$, $\gamma$, and $\delta$. The chains
formed by one of four kinds of Cu$^{2+}$ ions and relevant exchange
interactions\cite{Masuda_2005} within the system are shown schematically in
Fig.~\ref{fig:spins}(a).

The two-stage transition into a magnetically ordered state occurs at
$T_{c1}=24.6$~K and $T_{c2}=23.2$~K.\cite{Seki_2008} Neutron scattering
 and NMR experiments have revealed an incommensurate
magnetic structure in the magnetically ordered state
($T<T_{c1}$)~\cite{Masuda_2004, Gippius_2004, Kobayashi_2009}. The wave vector
of the incommensurate magnetic structure coincides with the chain direction
($b$-axis). The magnitude of the propagation vector at $T$ $<$ 17 K is almost
temperature independent and is equal to 0.827$\times$2$\pi$/$b$. The neutron
scattering experiments have shown that the magnetic moments neighboring along
the $a$--direction are antiparallel, whereas those neighboring along the
$c$--direction are coaligned. The intra-chain and inter-chain exchange
constants were determined from the analysis of the spin wave
spectra.\cite{Masuda_2005} The nearest neighbour exchange interaction is
ferromagnetic $J_1=-7.00$~meV, while the next nearest neighbour exchange
interaction is antiferromagnetic $J_2=3.75$~meV. The competition between these
interactions leads to incommensurate magnetic structure.
 The antiparallel
orientation of the magnetic moments of Cu$^{2+}$ of the neighbouring chains is
caused by the strong antiferromagnetic exchange interaction $J_3=3.4 $~meV. The
coupling of the Cu$^{2+}$ moments along the $c$-direction and the couplings
between the magnetic ions in different crystallographic positions are much
weaker \cite{Masuda_2005, Gippius_2004}. Thus, LiCu$_2$O$_2$ can be considered
as a quasi-two dimensional. The quasi-two-dimensional character of magnetic
interactions in LiCu$_2$O$_2$ compound was also proved by the resonant soft
X-ray magnetic scattering experiments~ \cite{Rusydi_2008, Huang_2008}.

Magnetic structure of LiCu$_2$O$_2$ at zero magnetic field was studied by
several groups by means of neutron diffraction experiments~\cite{Masuda_2004,
Seki_2008, Kobayashi_2009}. The authors of Ref.~\onlinecite{Masuda_2004} have
proposed the planar spiral spin structure with the spins confined to the
$(ab)$--plane. Polarized neutron scattering measurements of
Ref.\onlinecite{Seki_2008} have detected the spin component along the
$c$--direction, indicating the spiral magnetic structure in the $(bc)$--plane.
The authors of Ref.~\onlinecite{Kobayashi_2009}, alternatively, have proposed a
spiral spin structure confined to the (1,1,0) plane. It was attempted to
extract information about magnetic structure of LiCu$_2$O$_2$ from the studies
of electric polarization, which accompanies the magnetic
ordering~\cite{Park_2007,Seki_2008}. Unfortunately, at the moment, the nature
of this polarization  is not clear \cite{Moskvin_2009}, which did not permit to
obtain the definitive information about the zero-field magnetic structure from
this type of experiment.

 The magnetic structure realized in LiCu$_2$O$_2$
is much clearer in the magnetic fields above 3~T applied along $\vect{b}$ and
$\vect{c}$ directions. ESR and NMR studies of LiCu$_2$O$_2$ in the
low-temperature magnetically ordered phase ($T<T_{c2}$) have shown that the
planar spiral magnetic structure is formed in this
compound~\cite{Svistov_2009}. The magnetic moments located at the $\alpha$
($\beta$, $\gamma$ or $\delta$) position of the crystal unit cell with
coordinates $x$, $y$, $z$ (measured along
 the $a$, $b$ and $c$-- axes of
the crystal respectively) are defined as:

\begin{eqnarray}
\vectmu_\alpha=\mu \cdot \vect{l}_1 (-1)^{x/a} \cdot
\cos(k_{ic}\cdot y+\phi_\alpha)+ \nonumber\\
+\mu \cdot \vect{l}_2 (-1)^{x/a} \cdot \sin(k_{ic}\cdot
y+\phi_\alpha), \label{eqn:magnstr}
\end{eqnarray}

\noindent where $\vect{l}_1$ and $\vect{l}_2$ are the two mutually
perpendicular unit vectors, $\vect{k}_{ic}$ is the incommensurability vector
parallel to the chain direction ($b$-axis) and $\mu$ is the magnetic moment of
the Cu$^{2+}$ ion. The ordered magnetic moment per copper ion at $T\lesssim
10$~K was evaluated as $\mu=0.85$~$\mu_{B}$.\cite{Masuda_2004,Svistov_2009} The
phases $\phi_\alpha$, $\phi_\beta$, $\phi_\gamma$ and $\phi_\delta$ determine
the mutual orientation of the spins in the chains formed by the ions in the
different crystallographic positions. Their values can be extracted from the
NMR data\cite{Svistov_2009, Svistov_2010}.

\begin{figure}
\includegraphics[width=\figwidth,angle=0,clip]{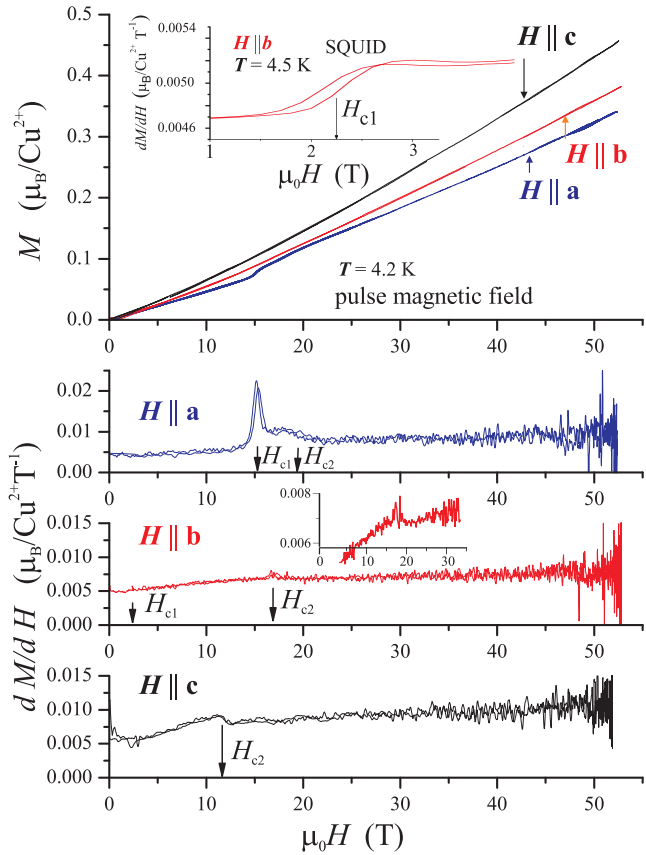}
\caption{(color online) Upper panel: Magnetization curves $M(H)$ at 4.2 K for
three field directions $\vect{H}\parallel \vect{a}, \vect{b}, \vect{c}$
measured in pulse magnetic fields of up to 52 T. The inset shows the field
dependence of $dM/dH$ at 4.5 K for $\vect{H} \parallel \vect{b}$ measured with
a SQUID magnetometer. Lower panels: Measured magnetic field dependences of
dM/dH for the principal field directions $\vect{H}\parallel \vect{a}, \vect{b},
\vect{c}$.  The inset shows the enlarged dependency in the range of transition
field $H_{c2}$ for $\vect{H} \parallel \vect{b}$. } \label{fig:M(H)}
\end{figure}

\begin{figure}
\includegraphics[width=\figwidth,angle=0,clip]{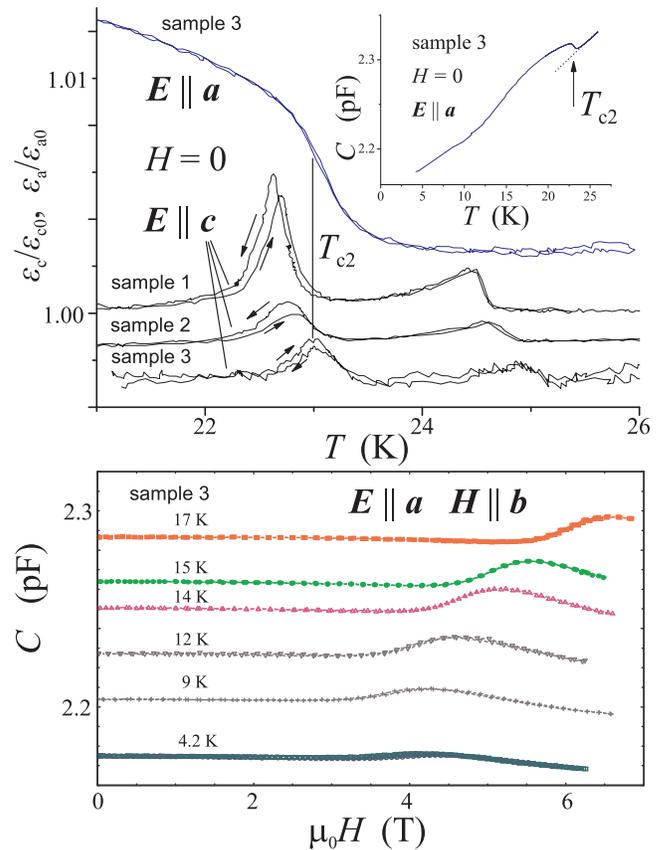}
\caption{(color online) Upper panel: Temperature dependences of dynamic
dielectric constants of three samples at zero magnetic field for alternating
electric field $\vect{E}\parallel \vect{a}, \vect{c}$. Dependences
$\varepsilon(T)$ presented in the main panel were obtained by subtraction of
the linear temperature drifts. Curves are shifted along the y-direction for
better presentation. Insert shows the raw temperature dependence of the
capacitance $C(T)$ for $\vect{E}\parallel \vect{a}$ at $H=0$ with the linear
extrapolation of $C(T)$  at $T>T_{c}$ to the range of $T \lesssim T_{c2}$ shown
by dotted line. Lower panel: The field dependences of the capacitance for
$\vect{E}\parallel \vect{a}$ and the magnetic field $\vect{H}\parallel
\vect{b}$ at different temperatures. The data in the lower panel are plotted
without shift along the y-axis.} \label{fig:epsilon}
\end{figure}

\section{Sample preparation and experimental details}
Single crystals of LiCu$_2$O$_2$ with the size of several cubic millimeters
were prepared by the solution in the melt method as described in
Ref.~\onlinecite{Svistov_2009}. The samples had a shape of a flat plate with
the developed $(ab)$--plane. The twinning structure of the samples was studied
by the optical polarization microscopy. Typical views of the $(ab)$--plane of
the samples are shown in Fig.~\ref{fig:domains}. The dark and bright areas on
the photos correspond to the domains with two different directions of the
$a$--axis. In most cases, the samples were twinned with characteristic domain
size of several microns (see Fig.~\ref{fig:domains}(c)). It was possible to
select the samples with large domains (see Fig.~\ref{fig:domains}(b)) and the
best samples were without twinning structure (Fig.~\ref{fig:domains}(a)).

The absence of twinning structure in the samples selected for the experiments
was confirmed by X-ray diffraction measurement(see lower panel of
Fig.~\ref{fig:domains}). The X-ray diffraction patterns were taken in
$\theta-2\theta$ geometry with use of CuK($\alpha$) irradiation. The untwinned
crystal produced clearly distinguishable diffraction patterns from the $(bc)$
and the $(ac)$--planes of the crystal (see lower panel of
Fig.~\ref{fig:domains}). The spectra obtained from twinned samples had a form
of superposition of these spectra. ESR measurements also confirm absence of
twinning in the samples used~\cite{Svistov_2009}.

Magnetization curves in static magnetic fields of up to 7~T were measured with
a commercial SQUID magnetometer (Quantum Design MPMS-XL7). High field
magnetization data were taken in magnetic fields of up to 52~T using a pulsed
magnet at KYOKUGEN in Osaka University.~\cite{Hagiwara_2006}

The dielectric constant $\varepsilon$ was obtained from the four-contact
capacitance measurements at 10~kHz on the capacitor formed by contact plates
attached to the sample. The naturally grown samples with the flat
$(ab)$--planes were used for the measurements of $\varepsilon_c$. Samples for
the measurements of $\varepsilon_a$ and $\varepsilon_b$ were properly cut from
the single crystals without twinning structure. Capacitor plates were formed
right on the sample surface  either by the sputtering of the silver film, or
with the silver paste.

 Different samples from the same batch were used for the
dielectric constant and magnetization measurements, these samples are further
denoted as "sample 1-5".

\section{Experimental results}

\begin{figure}
\includegraphics[width=\figwidth,angle=0,clip]{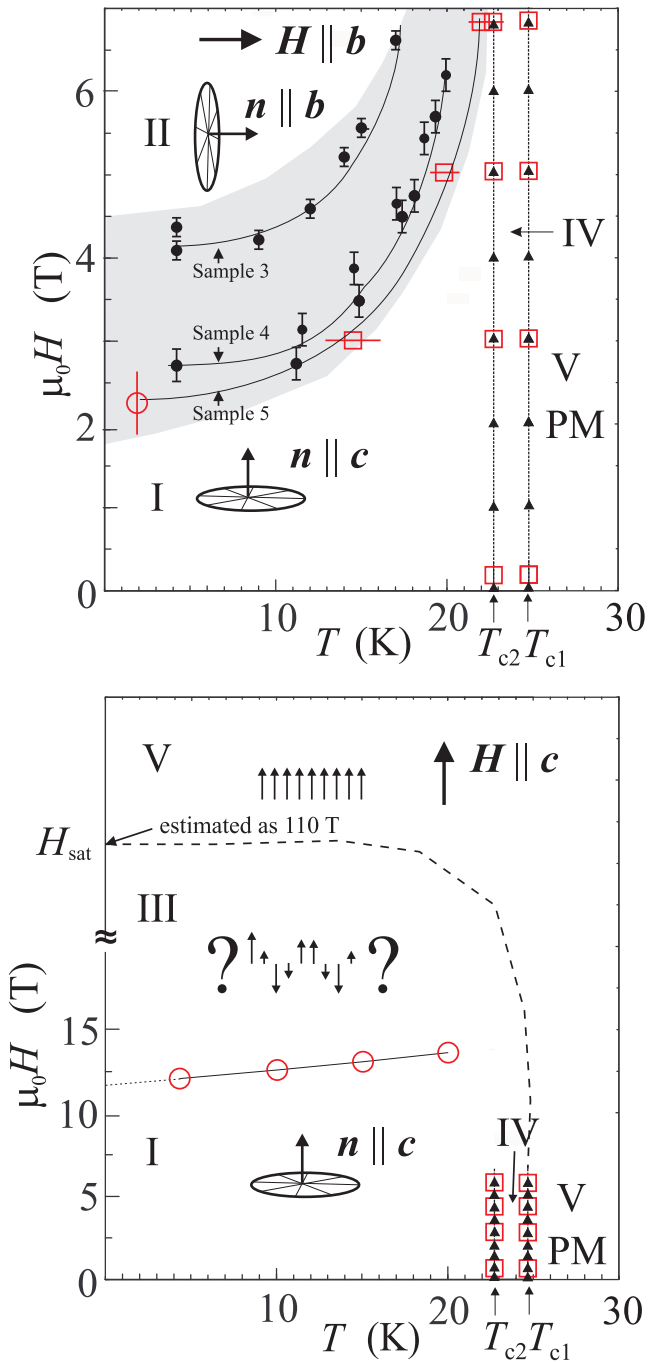}
\caption{(color online) Magnetic phase diagrams of LiCu$_2$O$_2$ compound for
$\vect{H}\parallel \vect{b}$ (upper panel) and $\vect{H}\parallel \vect{c}$
(lower panel). The different symbols in the figures correspond to the data from
different experimental methods. Solid triangles and circles show the phase
boundaries obtained from $\varepsilon_c(T)$ and $\varepsilon_a(H)$
measurements, respectively. Open red circles and squares show the phase
boundaries obtained from $M(H)$ and $M(T)$ measurements, respectively. The
solid lines are guides to the eye.  The grey shadowed area in the upper panel
marks sample-dependent position of the phase boundary between phases I and II.
High field spin structure III schematically shown in the phase diagram is
merely speculative. Dashed line is a schematic phase boundary between the
paramagnetic polarized and the ordered phases.} \label{fig:phasediagr}
\end{figure}

\subsection{Magnetization measurements at low fields}
The experimental curves $M(T)/\mu_0 H$ and their temperature derivatives for
three field directions $\vect{H}\parallel \vect{a}, \vect{b}, \vect{c}$ at
$\mu_{0}H=0.1$~T are shown in Figs. \ref{fig:M(T)} and \ref{fig:dmdt}. For all
three field directions, there is a broad maximum of $M(T)/\mu_0 H$ at $T=38$~K
which is a typical feature of low-dimensional antiferromagnets. In the
paramagnetic region, magnetic susceptibility for $\vect{H}\parallel \vect{c}$
exceeds that for $\vect{H}
\parallel \vect{a}, \vect{b}$, which is consistent with the anisotropy of the
$g$-tensor measured in ESR experiments~\cite{Vorotynov_1998} ($g_{a,b}= 2.0$;
$g_{c}=2.2$).  The magnetic susceptibilities are temperature independent
 for $T\ll T_N$. The absence of significant paramagnetic
contribution at low temperatures indicates low concentration of the
uncontrolled magnetic impurities.

The transitions to the magnetically ordered states at $T_{c1}$ and $T_{c2}$ are
marked by the change of the $M(T)$ slope. These inflection points are well
resolved on the temperature derivative of $M/\mu_0 H$ (Fig.~\ref{fig:dmdt}).
The $\frac{d}{dT}(M/\mu_0 H)$ curves are strongly anisotropic. For $\vect{H}
\parallel \vect{b}$ and $\vect{c}$, there are two anomalies corresponding to
two transitions at $T_{c1}$ and $T_{c2}$. The temperature dependence of
$\frac{d}{dT}(M/\mu_0 H)$ for $\vect{H}\parallel \vect{a}$ shows only one sharp
peak at the temperature near $T_{c2}$ in all studied field range. An additional
anomaly was observed on the $M(T)$ curves for $\vect{H}\parallel \vect{b}$ and
$\mu_0 H\gtrsim 3$~T (lower panel of Fig.~\ref{fig:M(T)}). At the transition
temperature $T_{c3}$, the slope of $M(T)$ changes abruptly. The value of
$T_{c3}$ increases with the increase of magnetic field and merges with $T_{c1}$
at $H\gtrsim 6$~T.

\subsection{High-field magnetization measurements}

The upper panel of Fig.~\ref{fig:M(H)} demonstrates magnetization $M(H)$ curves
measured in pulsed fields of up to 52~T for $\vect{H}\parallel \vect{a},
\vect{b}, \vect{c}$. These curves were obtained by integrating the $dM/dH$
curves presented in the lower three panels of Fig.~\ref{fig:M(H)}.

The $dM/dH$ curves show two anomalies which are marked as $H_{c1}$ and
$H_{c2}$. The anomaly at $H_{c1}$ was observed for $\vect{H}\parallel \vect{a}$
and for $\vect{H}\parallel\vect{b}$. It is characterized by the increase of the
$dM(H)/dH$, which may indicate a spin-reorientation transition. At the second
transition field $H_{c2}$, observed in all three principal orientations, the
field derivative of magnetization decreases.

The position of the critical field $H_{c1}$ for the $\vect{H}\parallel\vect{b}$
corresponds to the magnetic transition at the temperature $T_{c3}$ observed in
the temperature scan of magnetization. The transition field $H_{c1}$ is
strongly anisotropic: for the $\vect{H}\parallel\vect{a}$, $\mu_0H_{c1}=15$~T,
while for the $\vect{H}\parallel\vect{b}$ it varies from 2 to 5~T from sample
to sample and demonstrates a hysteresis. The differential susceptibility
$dM/dH$  increases stepwise at the transition  field $H_{c1}$ by $65\pm 5$ \%
for $\vect{H}\parallel \vect{a}$ and by $10\pm 3$\% for $\vect{H}\parallel
\vect{b}$.

The second transition field $H_{c2}$ is less anisotropic, its values are $\mu_0
H_{c2}=15.3, 16.7$ and $11.5$ T for the $\vect{H}\parallel
\vect{a},\vect{b},\vect{c}$, respectively. At this transition the differential
susceptibility  decreases stepwise by $25\pm3$ \%, $3\pm 1$\% and $11\pm 2$\%
for the $\vect{H}\parallel \vect{a},\vect{b},\vect{c}$, respectively.

\subsection{Dielectric constant measurements}
Magnetic and electric properties of LiCu$_2$O$_2$ are strongly
coupled~\cite{Park_2007,Seki_2008}. The magnetic transitions are
accompanied by anomalies of the dielectric constant.

The temperature and field dependences of the dielectric constant measured in
our experiments are shown in Fig.~\ref{fig:epsilon}. These curves were obtained
from the temperature dependences  of the capacitance. Linear drift obtained by
extrapolation of $C(T)$ at $T>T_c$ was subtracted for better presentation. The
example of the raw $C(T)$ curve and its high temperature linear extrapolation
is given on the insert to Fig.~\ref{fig:epsilon}. The temperature drift depends
on the quality of  capacitor plates contact with the sample. The strongest
temperature drifts were observed for samples prepared for measurements at
$E\parallel \vect{a},\vect{b}$. In these cases the samples were composed from
10 or more oriented parts fastened with glue.

There are clearly visible anomalies on $\varepsilon(T)$ at $E \parallel
\vect{a},\vect{c}$ while there is no anomaly at $E
\parallel \vect{b}$. The $\varepsilon_{c}(T)$ demonstrates two peaks at
$T_{c1}$ and $T_{c2}$ as shown in the upper panel of Fig.~\ref{fig:epsilon}.
These peaks correspond to the two successive transitions between the
paramagnetic phase and the ordered spiral one. The anomaly at $T_{c1}$ has no
hysteresis, while the $\varepsilon_c (T)$ shows hysteresis near $ T_{c2}$. The
positions of the peaks of $\varepsilon_c (T)$ varies from the sample to the
sample in the range of $24.7 \pm 0.2$~K and $22.7 \pm 0.2$~K for $T_{c1}$ and
$T_{c2}$, respectively. The peak value of the dielectric constant changes by
the factor of 10 depending on the sample. The $\varepsilon_a (T)$ shows only
one broad, step-like anomaly around $T_{c2}$. The positions of the peak of
$\varepsilon_c (T)$ and of the maximal slope of $\varepsilon_a (T)$ coincide
for the same sample. No anomalies on the temperature and field dependences of
$\varepsilon_{b}$ were observed.

One broad peak is observed  in the field dependence of capacitance $C$ of
LiCu$_2$O$_2$ sample for $\vect{H}\parallel \vect{b}$ and $\vect{E}\parallel
\vect{a}$ (lower panel of Fig.~\ref{fig:epsilon}), which corresponds to the
transition at $T_{c3}$ and $H_{c1}$  on the corresponding $M(T)$ and $M(H)$
dependences (see Figs.~\ref{fig:M(T)} and \ref{fig:M(H)}).

\subsection{Phase diagram}
The $H-T$ magnetic phase diagram of LiCu$_2$O$_2$ is presented in
Fig.~\ref{fig:phasediagr}. In total, it contains 5 different phases: (I) the
low-field phase, (II) the phase above $H_{c1}$ for $\vect{H}\parallel\vect{b}$,
(III) the phase above $H_{c2}$ for $\vect{H}\parallel\vect{c}$, (IV) the
intermediate ordered phase between $T_{c1}$ and $T_{c2}$, (V) the  polarized
paramagnetic phase.

The anomalies in $M(H)$, $M(T)$, $\varepsilon_a(T)$, $\varepsilon_a (H)$ which
indicate the transition from the phase I to the phase II on the phase diagram
have a hysteretic character.   The phase boundaries obtained for three
different samples for $\vect{H}\parallel\vect{b}$ are also shown on the phase
diagram. The boundary between phases I and II for $\vect{H}\parallel \vect{b}$
changes with the sample.

\section{Discussion}

To define the orientation of the spin plane in the space, it is convenient to
introduce the vector normal to the spin plane
$\vect{n}=[\vect{l}_1\times\vect{l}_2]$, where $\vect{l}_1$ and $\vect{l}_2$
are the order parameter components introduced in Eq.(\ref{eqn:magnstr}).
Orientation of the spin plane in a magnetic field is determined by the
competition of the anisotropy of magnetic susceptibility and the crystal
anisotropy. The Zeeman energy and energy of magnetic anisotropy for planar
spiral spin structure can be written as:

\begin{equation}
    E=-\frac{\chi_{\perp}}{2} H^2-\frac{\chi_{\parallel}-\chi_{\perp}}{2}(\vect{n}\cdot \vect{H})^2+a_1 n_x^2+a_2 n_y^2,
\end{equation}

\noindent where $\chi_{\parallel}$, $\chi_{\perp}$ are magnetic
susceptibilities for field applied parallel and perpendicular to vector
$\vect{n}$ (or, perpendicular and parallel to the spin plane, correspondingly).
The constants of anisotropy are positive and $a_{1}>a_{2}>0$; $x$ and $y$ axes
are aligned along the $a$ and $b$ axes, correspondingly. Such choice of
parameters ensures at zero field planar spiral structure with
$\vect{n}\parallel \vect{c}$, as it was suggested in
Ref.\onlinecite{Masuda_2004}.

As it  was argued above, the phase transition at $H_{c1}$ from the phase I to
the phase II is the reorientation transition. The observation of spin
reorientations for two field directions $\vect{H}\parallel \vect{a}$,
$\vect{H}\parallel \vect{b}$ is possible only for $\chi_{\parallel} >
\chi_{\perp}$: the spiral structure rotates in strong field to provide
$\vect{n}\parallel \vect{H}$. The critical fields for the
$\vect{H}\parallel\vect{a},\vect{b}$ can be obtained by minimizing of the
energy (Eq. 2) over the direction of the vector

\begin{eqnarray}
    \vect{H}\parallel\vect{a}:&&\nonumber\\
    H_{c1}&=&\sqrt{\frac{ 2a_{1}}{\chi_{\parallel}-\chi_{\perp}}}, \\
    \vect{H}\parallel\vect{b}:&&\nonumber\\
    H_{c1}&=&\sqrt{\frac{ 2a_{2}}{\chi_{\parallel}-\chi_{\perp}}},
\end{eqnarray}

The values of $\mu_0 H_{c1}$ for $\vect{H}
\parallel \vect{a}$ and $\vect{H}\parallel\vect{b}$ differ
essentially: 15 T and $\approx 1.8- 4.2$ T respectively (see phase diagram on
Fig. 7). Thus, we can conclude that $a_1/a_2\approx 12-70$. The anisotropy
parameters $a_1$ and $a_2$ define also the frequencies of electron spin
resonance (ESR). The spin reorientation transition at $\mu_0 H_{c1}\approx 16$
T for $\vect{H}\parallel \vect{a}$ was predicted from the ESR
data\cite{Svistov_2009}. Our high field M(H) experiments provide direct
experimental proofs for this predicted transition at $\mu_0 H_{c1}\approx 15$
T.

 The two stage
transition from the paramagnetic phase to the planar spiral phase is specific
for the magnetic systems with a strong "easy plane"  magnetic anisotropy for
the vector $\vect{n}$. \cite{Mochizuki_2010} This scenario is applicable to the
case of LiCu$_2$O$_2$ since  $a_1\gg a_2$. Thus, LiCu$_2$O$_2$ can be
approximately considered as a magnetic system with uniaxial anisotropy. For
these systems at $T_{c2}<T< T_{c1}$, only one component of the order parameter
of the spiral structure appears ($\mathbf{l_1}\parallel \mathbf{a}$, see
Eq.(\ref{eqn:magnstr}))  and the other is fluctuating, while both components
order at $T<T_{c2}$.

Peculiarity of the crystal $a$-direction is also confirmed by the magnetization
measurements. The temperature derivatives of the magnetization
(Fig.\ref{fig:dmdt}) for $\vect{H}\parallel\vect{b},\vect{c}$ are similar to
each other, and they demonstrate sharp anomalies both at $T_{c1}$ and at
$T_{c2}$. In the same time, the $dM/dT$ curve for $\vect{H}\parallel\vect{a}$
changes essentially only at $T_{c2}$.

In the high field range, the second phase transition was observed at $H_{c2}$.
This transition is accompanied by the decrease of the susceptibility and is
observed for all three principal field directions. This fact suggests the
exchange origin of this transition. At $H_{c2}$, a spiral structure in
individual chains probably changes to the other one. Such a transition from the
spiral to collinear spin-modulated phase was observed recently in the quasi-one
dimensional magnet with the same type of frustration LiCuVO$_4$.
\cite{Buettgen_2007} For the quasi-one dimensional systems, this unusual
magnetic phase was interpreted theoretically as a spin-density wave phase
\cite{ Hikihara_2008, Heidrich_2009, Sudan_2009}.

 For the one dimensional
frustrated chain with the exchange constants $J_1$ and $J_2$ of LiCu$_2$O$_2$
($J_1/J_2\approx -2$) corresponding transition is expected, according to
Ref.\onlinecite{Hikihara_2008}, at $H_{c2}\approx 0.2 H_{sat}$. The value of
$H_{sat}$ is not known experimentally, but it can be evaluated by extrapolating
the $M(H)$ dependences to the magnetization value equal to $gS\mu_B$ per
Cu$^{2+}$ ion. The linear extrapolation of $M(H)$ gives the upper estimate for
the saturation field $H_{sat}\approx$130~T. Taking into account the 15 percents
spin reduction in low field range, we can expect, that the saturation field
will be $\approx$ 15$\%$ less than the value derived from linear extrapolation:
$H_{sat}\approx$ 110 T. This results in the evaluation of the $H_{c2}$ value of
$\approx20-25$~T which is close to the values observed in experiment. The
theoretical consideration of the same 1D model in Ref.~\onlinecite{Sudan_2009}
predicts a transition to the spin-density wave phase at the value of magnetic
moment $0.12 M_{sat}$, which is also in the reasonable agreement with the
experimentally observed value $\approx 0.07 M_{sat}$ (see
Fig.~{\ref{fig:M(H)}).

 It is unclear how sensitive is the spin-density wave phase, predicted
theoretically for 1D system, to the interchain exchange interaction $J_3$,
which is essential for LiCu$_2$O$_2$ . The effect of the interchain exchange
interaction may be responsible for the difference between the results of the
model calculations and experimental values of $H_{c2}$. Thus  spin-density wave
phase can be a plausible candidate for magnetic phase realized in LiCu$_2$O$_2$
at $H>H_{c2}$. More definite conclusion requires further experimental study of
this phase.

Finally, it should be noted that the uniaxial anisotropy observed in
LiCu$_2$O$_2$ is unclear at the moment and still awaits theoretical
consideration.

\section{Conclusions}

The low field magnetic phase transition observed on the untwinned
LiCu${_2}$O$_2$ crystals  for the field aligned in the $(ab)$--plane of the
crystal can be described as a reorientation transition of the spiral structure.
The observed phase transitions are in agreement with the presence of the strong
easy $(bc)$--plane anisotropy for vector $\vect{n}$ of spiral structure in
LiCu$_2$O$_2$ \cite{Svistov_2009}. The origin of such strong anisotropy in
LiCu${_2}$O$_2$ at the moment is unknown. New high field magnetic transition
was observed for the field directed along all three principal axes of the
crystal. The critical field of this magnetic transition is close to the value
of transition field from a spiral to a spin density wave structure predicted
theoretically for one dimensional model with parameters of exchange
interactions $J_1$ and $J_2$ expected for LiCu$_2$O$_2$.

\acknowledgments
 This work was carried out
 under the Visiting Researcher Program of KYOKUGEN and
partly supported by Grants-in-Aid for Scientific Research (No.20340089), the
Global COE Program (Core Research and Engineering of Advanced
Materials-Interdisciplinary Education Center for Materials Science) (No. G10)
from the MEXT, Japan, by the Grants 12-02-00557-à, 10-02-01105-a of the Russian
Foundation for Basic Research, and Program of Russian Scientific Schools.


\begin{thebibliography}{35}
\bibitem{Chubukov_1991} A.~V.~Chubukov, Phys. Rev. B {\bf{44}}, 4693
(1991).
\bibitem{Kolezhuk_2000} A.~K.~Kolezhuk, Phys. Rev. B {\bf{62}}, R6057,
(2000).
\bibitem{Kolezhuk_2005}A.~K.~Kolezhuk and T.~Vekua, Phys. Rev. B {\bf{72}}, 094424
(2005).
\bibitem{Vekua_2007}T.~Vekua, A.~Honecker, H.-J.~Mikeska, and F.~Heidrich-Meisner,  Phys. Rev. B
{\bf{76}}, 174420, (2007).
\bibitem{Dmitriev_2008} D.~V.~Dmitriev and V.~Yu.~Krivnov, Phys. Rev. B {\bf{77}}, 024401
(2008).
\bibitem{Heidrich_2006}F.~Heidrich-Meisner, A.~Honecker and T.~Vekua, Phys. Rev. B {\bf{74}},
020403(R), (2006).
\bibitem{Hikihara_2008} T.~Hikihara, L.~Kecke, T.~Momoi and A.~Furusaki Phys. Rev. B {\bf{78}}, 144404 (2008).
\bibitem{Sudan_2009} J.~Sudan, A.~L\"{u}scher, and A.~M.~L\"{a}uchli, Phys. Rev. B {\bf 80},
140402(R) (2009).
\bibitem{Heidrich_2009} F.~Heidrich--Meisner, I.~P.~McCulloch, and A.~K.~Kolezhuk,
Phys. Rev. B {\bf 80}, 144417 (2009).
\bibitem{Furukawa_2008} S.~Furukawa, M.~Sato, Y.~Saiga, and S.~Onoda,
 J. Phys. Soc. Jpn. {\bf 77},123712 (2008).
\bibitem{Furukawa_2010} S.~Furukawa, M.~Sato and S.~Onoda, Phys. Rev. Lett.
{\bf{105}}, 257205, (2010).
\bibitem{Zhitomirsky_2010} M.~E.~Zhitomirsky and H.~Tsunetsugu, Europhys. Lett. {\bf92}, 37001 (2010).
\bibitem{Nishimoto_2010} S.~Nishimoto, S.~-L.~Drechsler, R.~Kuzian  and J.~van~den~Brink cond-mat, arXiv:1005.5500
(2010)
\bibitem{Sato_2011}M.~Sato, S.~Furukawa,S.~Onoda and A.~Furusaki, Mod. Phys.
Lett. B {\bf{25}},901, (2011).
\bibitem{Enderle_2005} M.~Enderle, C.~Mukherjee, B.~F\"ak, R.~K.~Kremer, J.~-M.~Broto, H.~Rosner, S.~-L.~Drechsler,
J.~Richter, J.~Malek, A.~Prokofiev, W.~Assmus, S.~Pujol, J.~-L.~Raggazzoni,
H.~Rakoto, M.~Rheinst\"adter, H.~M.~R{\o}nnow, Europhys. Lett. {\bf 70}, 237
(2005).
\bibitem{Hase_2004} M.~Hase, H.~Kuroe, K.~Ozawa, O.~Suzuki, H.~Kitazawa, G.~Kido, and T.~Sekine, Phys. Rev. B {\bf 70} 104426 (2004).
\bibitem{Drechsler_2007} S.~-L.~Drechsler, O.~Volkova, A.~N.~Vasiliev, N.~Tristan, J.~Richter, M.~Schmitt, H.~Rosner, J.~M\'alek, R.~Klingeler, A.~A.~Zvyagin, and B.~B\"uchner, Phys. Rev. Lett. {\bf 98} 077202 (2007).
\bibitem{Banks_2007} M.~G.~Banks, F.~Heidrich-Meisner, A.~Honnecker,
H.~Rakoto, J.~-M.~Broto, R.~K.~Kremer, J. Phys.: Condens. Matter {\bf 19},
145227 (2007).
\bibitem{Masuda_2005} T.~Masuda, A.~Zheludev, B.~Roessli, A.~Bush, M.~Markina, and A.~Vasiliev, Phys. Rev. Â {\bf 72}, 014405 (2005).
\bibitem{Svistov_2009} L.~E.~Svistov, L.~A.~Prozorova, A.~M.~Farutin, A.~A.~Gippius, K.~S.~Okhotnikov,
A.~A.~Bush, K.~E.~Kamentsev, and \'E.~A.~Tishchenko, JETP {\bf 108}, 1000
(2009).
\bibitem{Berg_1991} R.~Berger, A.~Meetsma, S.~v.~Smaalen, J. Less-Common Met.
$\bf{175}$, 119, (1991).
\bibitem{Masuda_2004}T.~Masuda, A.~Zheludev, A.~Bush, M.~Markina, and A.~Vasiliev, Phys. Rev. Lett. {\bf 92}, 177201 (2004).
\bibitem{Seki_2008} S.~Seki, Y.~Yamasaki, M.~Soda, M.~Matsuura, K.~Hirota, and Y.~Tokura, Phys. Rev. Lett. {\bf 100}, 127201 (2008).
\bibitem{Gippius_2004} A.~A.~Gippius, E.~N.~Morozova, A.~S.~Moskvin, A.~V.~Zalessky, A.~A.~Bush, M.~Baenitz, H.~Rosner, and S.~-L.~Drechsler, Phys. Rev. Â {\bf 70}, 020406 (2004).
\bibitem{Kobayashi_2009} Y.~Kobayashi, K.~Sato, Y.~Yasui, T.~Moyoshi, M.~Sato, and K.~Kakurai, J. Phys. Soc. Jpn. {\bf 78}, 084721 (2009).
\bibitem{Rusydi_2008} A.~Rusydi, I.~Mahns, S.~M\"uller, M.~R\"ubhausen, S.~Park, Y.~J.~Choi, C.~L.~Zhang,
S.~-W. Cheong, S.~Smadici, P.~Abbamonte, M.~v.~Zimmermann, and G.~A.~Sawatzky,
Appl. Phys. Lett. {\bf 92}, 262506 (2008).
\bibitem{Huang_2008} S.~W.~Huanga, D.~J.~Huanga, J.~Okamoto, W.~B.~Wu, C.~T.~Chen, K.~W.~Yeh, C.~L.~Chen, M.~K.~Wu,
H.~C.~Hsu, F.~C.~Chou,, Sol. State Com. {\bf 147}, 234 (2008).
\bibitem{Park_2007} S.~Park, Y.~J.~Choi, C.~L.~Zhang, and S.~-W.~Cheong, Phys. Rev. Lett. {\bf 98} 057601 (2007).
\bibitem{Moskvin_2009}A.~S.~Moskvin, Yu.~D.~Panov, and S.~-L.~Drechsler, Phys. Rev. B
{\bf 79}, 104112 (2009).
\bibitem{Svistov_2010} L.~E.~Svistov, L.~A.~Prozorova, A.~A.~Bush, K.~E.~Kamentsev,
J. Phys.: Conf. Series {\bf 200}, 022062 (2010).
\bibitem{Hagiwara_2006} M.~Hagiwara, S.~Kimura, H.~Yashiro, S.~Yoshii and K.~Kindo,
 J. Phys. Conf. Ser. {\bf 51}, 647 (2006).
\bibitem{Vorotynov_1998}A.~M.~Vorotynov, A.~I.~Pankrats, G.~A.~Petrakovskii, K.~A.~Sablina
W.~Paszkowicz and H.~Szymczak, JETP {\bf 86}, 1020 (1998).
\bibitem{Buettgen_2007} N.~B\"{u}ttgen, H.~-A.~Krug~von~Nidda, L.~E.~Svistov, L.~A.~Prozorova, A.~Prokofiev,
W.~Assmus, Phys. Rev. B {\bf{76}}, 014440 (2007).
\bibitem{Mochizuki_2010} M.~Mochizuki and N.~Furukawa, Phys. Rev. Lett. {\bf 105}, 187601
(2010).


\end{thebibliography}
\end{document}